\documentstyle[12pt]{article}
\begin{document}

\vskip 1truecm
\rightline{Preprint PUPT-1557 (1995)}
\rightline{ e-Print Archive: hep-ph/9508405
}
\bigskip
\centerline{\Large Fermion Determinant and the Sphaleron Bound}
\medskip
\centerline{\Large Guy D. Moore\footnote{e-mail:
guymoore@puhep1.princeton.edu}
}
\medskip

\centerline{\it Princeton University}
\centerline{\it Joseph Henry Laboratories, PO Box 708}
\centerline{\it Princeton, NJ 08544, USA}
\medskip

\centerline{\bf Abstract}
We investigate analytically the fermionic fluctuation determinant at finite
temperatures in the minimal
standard model, including all operators up to dimension 6 and all
contributions to the effective
potential to all orders in the high $T$ expansion, to 1 loop.  We apply
the results to find corrections to the Sphaleron erasure rate in the
broken phase.  We conclude that the analytic treatment
of fermions is very reliable, and
that there is a great deal of baryon erasure
after the phase transition for any physical Higgs mass.

\smallskip

\section{Introduction}

It has been known since the pioneering work of Kirzhnits and Linde \cite{Linde}
and Dolan and Jackiw \cite{Jackiw} that, at high temperature, the thermal
populations of massive particles exert a symmetry restoring force on the
Higgs condensate, so that at sufficiently high temperaure ($T \sim m_H/g$)
the Higgs field loses its condensate and electroweak symmetry is restored.
Under these circumstances baryon number is readily violated
\cite{Rubakov,McLerran}.  As temperature falls there is a symmetry breaking
phase transition, which is first order for small $m_H$.  Presumably the
universe underwent such a phase transition shortly after the Big Bang, and
it is believed likely that the baryon asymmetry of the universe was
generated at this time.

Equilibrium thermodynamic information important to the physics of
this epoch, such as the effective potential
and the free energy of certain saddlepoint solutions,
can be computed in the Matsubara (imaginary time) formulism.
The particular calculation we will be interested in is the Sphaleron rate
in the broken phase shortly after the phase transition.  This was first
investigated at tree level in \cite{Klink}, and one loop corrections have
been computed within a high temperature approximation in
\cite{Carson,Junker}.  This approximation assumes that the contributions of
nonzero Matsubara frequencies can be absorbed into shifts in the parameters
of the zero Matsubara frequency modes, which then constitute a ``dimensionally
reduced'' three dimensional theory\cite{Landsman}.
The same idea forms the basis for
an extensive investigation of the effective potential and the strength
of the phase transition being carried out by Farakos, Kajantie, Laine,
Rummukainen, and Shaposhnikov (FKLRS) \cite{FKRS,FKLRS}.

Why is it necessary to include 1 loop corrections?
Normally, one loop corrections
modify tree level results by only a few percent, but there are exceptions.
The finite temperature, 1 loop correction to the Higgs mass is significant
because, although the correction
is parametrically order $g^2$, the mass is a dimensionful
parameter, and the importance of the thermal effect is enhanced by $T^2/m^2$,
which can be large.  The phase transition temperature is
the point where the thermal, 1 loop effect and the tree level effect are
almost equal.  Finite temperature calculations also have potential infrared
divergences from loops containing zero Matsubara frequency bosons,
and it is not clear that perturbation theory always works; the object of
the dimensional reduction program is to separate this problem from those
effects which can be treated perturbatively more reliably.  Finally,
loop corrections can also be important when a coupling
is unnaturally small; for instance, the Higgs self coupling $\lambda$ recieves
a 1-loop correction which goes as $g_t^4$ ($g_t$ is the Yukawa coupling of
the top quark.  In this paper we will always take $g_t=1$, corresponding to
$m_t \simeq 174$GeV).  This correction does not depend on $\lambda$, and if
$\lambda$ happens to be small
then the correction can be very important.  This is the case in the
standard model whith a light Higgs boson and a heavy top quark.  It is
quite necessary, then, to consistently use 1 loop relationships when
describing anything directly related to $\lambda$, such as the physical
Higgs mass and the (finite and zero temperature) effective potential.  In
this case, however, two loop effects are still expected to make a small
perturbative correction to the one loop effects, so the one loop calculation
should be reliable.

With this in mind, Diakonov, Polyakov, Sieber, Schaldach, and Goeke
(DPSSG) have made a direct numerical evaluation of the fermionic fluctuation
determinant and have concluded on its basis that, when the
top quark is heavy, fermion fluctuations make the Sphaleron rate very
different from the high temperature estimate \cite{DPSSG}.  This brings the
dimensional reduction program into question, and bears further investigation.
Here we give a one loop, perturbative treatment of
the fermions in the presence of zero Matsubara frequency background fields.
The free energy of the fermions can naturally be expressed in terms of
operators made up of zero Matsubara frequency bosonic fields and their
derivatives.  In section II we compute these, including induced masses
(dimension 2), corrections to couplings and wave function renormalizations
(dimension 4), and nonrenormalizeable dimension 6 operators.  The expansion
in operator dimension appears to be very well behaved when the background
bosonic fields are slowly varying.
In section III we apply these results to a computation of the
Sphaleron energy.  In section IV we use this calculation, along with 1-loop
relationships between the couplings and the physical Higgs mass and a
calculation of the bubble nucleation rate which determines the temperature
at which the phase transition occurs, to investigate the erasure of
baryons after the phase transition is complete.  In the last section we
draw conclusions.  Some technical details are relegated to an appendix.

\section{Integration over fermions}

We will work in SU(2) Higgs theory coupled to all the fermions of the
standard model.  Neglecting hypercharge significantly simplifies the
picture whithout profoundly changing it, because the Weinberg angle is
small and because the fields which make up the Sphaleron are almost
exclusively weak isospin, and not hypercharge, fields.
(It is known that, at tree level,
including hypercharge with the physical Weinberg angle only changes the
Sphaleron energy by about 1\% \cite{Klink}).  The Lagrangian, suppressing
summations on generations and colors, is
\begin{eqnarray}
{\cal L} & = & {\cal L}_{b} + {\cal L}_{f}
\nonumber \\
{\cal L}_{b} & = &
\frac{1}{4} F_{\mu \nu}^a F_{\mu \nu}^a + (D_\mu \Phi)^{\dag} D_\mu \Phi
- m_0^2 \Phi^{\dag} \Phi + \lambda (\Phi^{\dag} \Phi)^2
\nonumber \\
{\cal L}_{f} & = &  \overline{\psi}_L \gamma_\mu D_\mu \psi_L
+ \overline{\psi}_R \gamma_\mu \partial_\mu \psi_R + g_t (\overline{\psi}_L
\Phi \psi_R + \overline{\psi}_R \Phi^{\dag} \psi_L ) \, .
\nonumber
\end{eqnarray}
(We use a Euclidean metric
$g_{\mu \nu} = \delta_{\mu \nu}$ and Euclidean $\gamma$ matricies which
satisfy the algebra $ \gamma_\mu \gamma_\nu + \gamma_\nu \gamma_\mu =
2 \delta_{\mu \nu}$.)
The fermion kinetic terms apply for every doublet and singlet, but the
mass term only applies for the top quark.  Actually the mass term as written
is for a bottom, not top, type quark.  This is for
notational simplicity only; to make the top quark massive one should
systematically replace $\Phi$ with $i \tau_2 \Phi^*$ in all that
follows; our conclusions are completely unchanged.

In the 1 loop approximation we ignore the fermions' coupling to nonzero
Matsubara frequency excitations of the bosonic fields.  The path integral
over the fermions is gaussian; its contribution to the partition function
is Det$H$, and the corresponding contribution to the effective action
is $-$Tr ln$H$, where $H$, written as a matrix acting
on  $[\psi_L^\alpha \; \; \psi_R]^T$, is

\begin{equation}
H = \left[ \begin{array}{cc} \gamma_\mu D^\mu_{\beta \alpha}
 & g_t \Phi_\beta \\
     g_t \Phi^{\dag}_\alpha & \gamma_\mu \partial^\mu \end{array} \right]
= \left[ \begin{array}{cc}   \gamma_{\mu} \partial^{\mu}
      \delta_{\beta \alpha} & 0 \\ 0 & \gamma_{\mu} \partial^{\mu}
	\end{array} \right] + \left[ \begin{array}{cc}
	\frac{ig_w}{2} \gamma_{\mu} A^{\mu}_{a} \tau^{a}_{\beta \alpha} &
	g_t \Phi_{\beta} \\ g_t \Phi^{\dag}_{\alpha} & 0 \end{array} \right]
\equiv H_0 + H_I \, ,
\label{Hamiltonian}
\end{equation}
where $\alpha$ and $\beta$ are $SU(2)$ indicies.

Our analysis will be based on an expansion of $-{\rm Tr} \ln H$ in
$H_I$.  To illustrate the idea of expanding the log,
consider a simplified example
in which the gauge field is everywhere zero.  In this case we have a Dirac
fermion with a (spatially varying) mass, $H = \gamma^\mu \partial_\mu - m$,
where $m^2 = g_t^2 \Phi^{\dag} \Phi$.
We assume that $m^2$ is smaller than the lowest eigenvalue of
$- \partial^2$, in which case it is legitimate to expand the log.
This will generally be the case, as the lowest eigenvalue of $-\partial^2$
is set by the square of the lowest possible Matsubara frequency, which for
a fermion is $(\pi T)^2$.  The log becomes
\begin{equation}
-{\rm Tr} \, \ln ( \gamma^\mu \partial_\mu - m ) =
 -{\rm Tr}\, \ln \gamma^\mu \partial_\mu
 - \sum_{n=1}^{\infty} \frac{(-1)^{n+1}}{n} {\rm Tr} \left(
\frac{m}{\gamma_\mu \partial_\mu} \right)^n
\end{equation}
where as usual $1/(\gamma^\mu \partial_\mu)$ is defined as $\gamma^\mu
\partial_\mu / \partial^2$.  We get Feynman diagrams by the usual trick of
inserting complete sets of states and Fourier transforming.  The first term
in the sum is a divergent vacuum energy and should be removed.  All terms with
odd powers of $m$ vanish when we take the trace on Dirac indicies.  When
the mass is position independent, the resulting terms are
\begin{equation}
\int_0^{T^{-1}} \! \! dx_0 \int d^3 x
\sum_{n=1}^{\infty} \, (-)^{n} \frac{4m^{2n}}{2n} T \sum_{k_0}
\int \frac{d^3 k}{(2\pi)^3} \frac{1}{(k^2 + k_0^2)^n} \, .
\label{massexpan}
 \end{equation}
We have used the shorthand $\sum_{k_0}$ to mean a sum in which $k_0$ takes
on odd integer multiples of $\pi T$; we will use this shorthand throughout.
Also, $k^2$ means $\vec{k}^2$ and $k$ means $\sqrt{\vec{k}^2}$.
The $(-)^n$ arises because when we Fourier transform $\partial \rightarrow
ik$.  The $4$ is from the Dirac trace, and $2n$ is the symmetry factor
of the diagram; we saw how it arises in the expansion of the log above.
To get the free energy density we drop the space integral.

The first two integrals in the series are ultraviolet divergent and must be
performed with some care.  We have done so in the $\overline{\rm MS}$ scheme
by first performing the sum on Matsubara modes and conducting the spatial
integrals in $3-2 \epsilon$ dimensions.
The higher order terms are convergent and may be performed directly by doing
the integral over $d^3k$ first.  Performing the
integrals, the free energy per unit volume is
\begin{equation}
m^2 \frac{T^2}{12} + m^4 \frac{2\gamma_E - 2 \ln \pi + \ln (\mu^2/T^2)}
{16\pi^2} + \sum_{n=3}^{\infty} (-)^n \frac{4 m^{2n}}{2n}
\frac{ (1-2^{3-2n}) \zeta(2n-3) \Gamma(n-\frac{3}{2})}{4 \pi^{2n-2} T^{2n-4}
\Gamma(\frac{1}{2}) \Gamma(n)}
\label{hiTexpan}
\end{equation}
with $\mu$ the usual $\overline{\rm MS}$ renormalization point.
This is the 1 loop fermionic contribution to the finite temperature effective
potential, to all orders in the high temperature expansion\footnote{An
integral expression for the fermionic contribution was first found by
Dolan and Jackiw \cite{Jackiw}, who also found the first two terms in the
expansion given here.  What we have done is found the complete Taylor series
for the known integral expression.}.
Note that, as expected, the sum is a Taylor
series in $m^2$ with radius of convergence $(\pi T)^2$.  By $m_t$ we mean
the thermal top quark mass given by $m_t^2(T) = g_t^2 \Phi^{\dag} \Phi (T)$,
not the vacuum top quark mass $m_t^2(0) = g_t^2 \Phi^{\dag} \Phi (T = 0)$.
Near the phase transition temperature the difference is quite substantial.
In the simplest approximation $\Phi^{\dag} \Phi (T) = \Phi^{\dag} \Phi(T=0)
(1-T^2/T_c^2)$, but near $T_c$ it is necessary to treat the influence of
infrared bosons more carefully.  As we will see in Sec. IV, at the
temperature of interest the Higgs VEV
$\nu \equiv \sqrt{\Phi^{\dag} \Phi/2} \leq T$ and
for $g_t=1$ we find $(m/\pi T)^2 \leq 1/(2 \pi^2)$, so the convergence of
the series is excellent.  Of course, well below the phase transition
temperature it is no longer true that $m_t << \pi T$ and our approximation
scheme will fail.  However, at these temperatures the Sphaleron energy
is so large that the baryon erasure rate is utterly negligible, so our
approximation scheme should apply in the interesting range of temperatures.

Another way of understanding the convergence of the power series is to think
of the different Matsubara frequency contributions of the top quarks as
distinct, very massive species of a three dimensional theory, with masses
given by the Matsubara frequencies $m_M^2 = ( (2n+1) \pi T)^2$; we are
then expanding in the ratio of $g_t^2 \Phi^{\dag} \Phi$ (and eventually
in products of other infrared bosonic fields like $A_i$, and their
derivatives) to $m_M^2$.  Such an expansion has been considered in a theory
with only fermions and a scalar in \cite{Baacke1,Baacke2}, where it
is also concluded that such an expansion is very accurate.  An expansion
like the one used here is not justified for bosons because
symmetric boundary conditions in time make the lowest
eigenvalue of the operator $\partial^2$ is zero; or equivalently the
lowest Matsubara frequency is zero, so we cannot expand in the ratio of
a field value to this Matsubara frequency.  (However we could expand all
the other Matsubara frequencies in the way described here, see \cite{FKRS}.)

For simplicity of notation, in the remainder of the paper we write
\begin{equation}
T \sum_{k_0} \int \frac{d^3k}{(2\pi)^3}
\frac{1}{(k^2 + k_0^2)^2}   =
\frac{2 \gamma_E - 2 \ln \pi + \ln (\mu^2 / T^2)}{16 \pi^2} \equiv D4
\label{D4def}
\end{equation}
and
\begin{equation}
 T \sum_{k_0} \int \frac{d^3k}{(2\pi)^3}
\frac{1}{(k^2 + k_0^2)^N}
 = \frac{(1-2^{3-2N}) \zeta(2N-3)
\Gamma(N-\frac{3}{2})}{4\pi^{2N-2} T^{2N-4} \Gamma(\frac{1}{2}) \Gamma(N)}
 \equiv  D2N \, .
\end{equation}
in particular,
\begin{equation}
D6 = \frac{7 \zeta(3)}{128 \pi^4 T^2} \, .
\end{equation}

When the mass is spatially varying the first term in the
sum in Eq.(\ref{massexpan}) becomes
\begin{equation}
\int dx_0
\frac{-1}{2} \int \frac{d^3p}{(2 \pi)^3} m(p) m(-p)
 T \sum_{k_0}
 \int \frac{d^3 k}{(2\pi)^3}
\frac{ {\rm tr} ( \gamma^\mu \gamma^\nu ) k_\mu (p+k)_\nu}
{(k^2 + k_0^2) ((p+k)^2 + k_0^2)} \, .
\label{derivexpan}
\end{equation}
If we assume that $p^2 < k^2 + k_0^2$, which means that the mass
varies on a scale
large compared to $1/\pi T$, then we may expand the denominator in a geometric
series and extract a power series in $p^2$.  We can then Fourier transform
to position space to express the result as a derivative expansion for $m$.
The resulting free energy is
\begin{equation}
\int d^3x  \left( \frac{T^2}{12} m^2 + D4 \, (\vec{\nabla} m)^2
- \frac{D6}{3} (\nabla^2 m)^2
 + \frac{D8}{10} ( \vec{\nabla} \nabla^2 m)^2
\ldots \right) \, .
\label{asymptotic}
\end{equation}
In realistic cases the Fourier transform of $m$ lies primarily at small
values of $p$ but has a rapidly decaying exponential tail which goes
above $p = \pi T$.  In this case the series will be asymptotic and its
reliability will depend on how rapidly the exponential tail falls off.
For the Sphaleron at $T \simeq T_c$ we expect the exponential tail of
the Fourier transform of the gauge and Higgs field configurations to
fall roughly as $\exp(-p/m_W (T))$; at $T \simeq T_c$,
$m_W(T) = g_w \nu(T) /2$ is $<< \pi T$ and the convergence
of the (asymptotic) derivative expansion should be excellent.  Of course
we will check this by explicit calculation.

For the more general $H$ the lowest nonvanishing term will be
\begin{eqnarray}
\frac{1}{2} {\rm Tr} \frac{H_I}{H_0^2} H^0 \frac{H_I}{H_0^2} H_0
\Rightarrow \frac{1}{2} \int \frac{d^3 p}{(2 \pi)^3}
T \sum_{k_0} \int \frac{d^3 k}{(2 \pi)^3}
(\Phi^{\dag}_{\alpha}(p) \Phi_{\alpha}(-p))
\frac{i k_{\mu} i (k+p)_{\nu} {\rm tr}(\gamma^{\mu} \gamma^{\nu})}{(k^2+k_0^2)
( (k+p)^2 + k_0^2)}
\nonumber \\
+ \frac{1}{2} \int \frac{d^3 p}{(2 \pi)^3} T \sum_{k_0} \int \frac{k^3}
{(2 \pi)^3} \left( \frac{g_w^2 \delta_{ab}}{2} i A_{\mu}^a(p)
i A_{\nu}^b(-p) \right)
\frac{i k_{\alpha} i (p+k)_{\beta} {\rm tr} ( \frac{1-\gamma^5}{2}
\gamma^{\alpha} \gamma^{\mu} \gamma^{\beta} \gamma^{\nu})}{(k^2 + k_0^2)
( (k+p)^2 + k_0^2)}
\end{eqnarray}
which corresponds to the Feynman diagrams illustrated in Fig. \ref{one}.
Higher order contributions can be gotten similarly by going to higher
powers in $H_I$.

The contribution of fermionic fluctuations to dimension two and four operators
in $SU(2) \times U(1)$ Higgs theory in $\overline{\rm MS}$
with realistic couplings
has recently been worked out by FKLRS \cite{FKLRS}.  We independently
performed the calculation in $SU(2)$ Higgs theory; our results concur.
We have also extended the calculation to find all nonrenormalizeable operators
induced by fermions at dimension 6.  Our results follow.

The dimension two operators induced by fermions change the masses of the
Higgs and $A_0$ fields.  The induced terms are
\begin{equation}
\frac{N_c g_t^2 T^2}{12} \Phi^{\dag} \Phi
+ \frac{N_d g_w^2 T^2}{24} A_0^2 \, .
\end{equation}
The first term is largely responsible for the restoration of symmetry at
high temperature.  The second is the familiar Debeye screening mass.
(Here and throughout $N_c = 3$ is the number of colors and $N_d = 12$ is
the number of left handed fermion doublets.)

At dimension four, fermions introduce the following corrections to couplings
and wave functions:
\begin{eqnarray}
N_c g_t^4 D4 \, (\Phi^{\dag} \Phi)^2
+ N_c g_t^2 D4 \, (D_i \Phi)^{\dag} D_i \Phi
+ \frac{N_c g_w^2 g_t^2}{4} (D4 - \frac{1}{8\pi^2}) A_0^2 \Phi^{\dag} \Phi
\nonumber \\
+ \frac{N_d g_w^2 D4}{12} F^a_{ij}F^a_{ij}
+ \frac{N_d g_w^2}{6}(D4 - \frac{1}{16\pi^2}) (D_i A_0)^a (D_i A_0)^a
- \frac{N_d g_w^4}{192 \pi^2} (A_0^2)^2 \, .
\label{dim4contrib}
\end{eqnarray}
The coefficients of $D4$ give the fermionic contributions to the one loop
beta functions and anomalous dimensions for the bosonic fields in $SU(2)$
Higgs theory, and agree with well known results.  If the dimensional
reduction scheme is justified then these terms should give an accurate
approximation of the fluctuation determinant.  To check this it is necessary
to go one order higher and see if the contributions of dimension 6
operators are small.

At dimension six, fermions contribute the following terms to the effective
potential,
\begin{equation}
-\frac{2 N_c g_t^6 D6}{3}(\Phi^{\dag} \Phi)^3
+ \frac{N_c g_t^2 g_w^4 D6}{16} (A_0^2)^2 \Phi^{\dag} \Phi \, ,
\end{equation}
the following Higgs derivative terms,
\begin{eqnarray}
- \frac{ 4 N_c g_t^4 D6}{3} (\Phi^{\dag} \Phi) (D_i \Phi)^{\dag} D_i \Phi
- \frac{ 2 N_c g_t^4 D6}{3} (\Phi^{\dag} D_i \Phi) (D_i \Phi)^{\dag} \Phi
\nonumber \\
- \frac{ N_c g_t^4 D6}{3} \partial_i(\Phi^{\dag} \Phi) \partial_i
       (\Phi^{\dag} \Phi)
-\frac{N_c g_t^2 D6}{3} (D^2 \Phi)^{\dag} D^2 \Phi \, ,
\label{Higgsderivs}
\end{eqnarray}
the following mixed derivative terms,
\begin{equation}
\frac{N_c g_w^2 g_t^2 D6}{6} A_0^2 (D_i \Phi)^{\dag} D_i \Phi
- \frac{N_c g_w^2 g_t^2 D6}{12} (D_i A_0)^a (D_i A_0)^a \Phi^{\dag} \Phi \, ,
\end{equation}
the following $A_0$ derivative terms,
\begin{equation}
\frac{N_d g_w^4 D6}{24} \partial_i (A_0^2) \partial_i (A_0^2)
- \frac{N_d g_w^2 D6}{30} (D^2 A_0)^a (D^2 A_0)^a \, ,
\end{equation}
the following mixed terms,
\begin{equation}
- \frac{N_c g_t^2 g_w^2 D6}{8} F^a_{ij} F^a_{ij} \Phi^{\dag} \Phi
- \frac{N_d g_w^4 D6}{24} A_0^2 F^a_{ij} F^a_{ij}
+ \frac{N_d g_w^4 D6}{8} A_0^a F^a_{ij} A_0^b F^b_{ij} \, ,
\end{equation}
and the following gauge field terms,
\begin{equation}
-\frac{N_d g_w^3 D6}{180} f_{abc} F^a_{ij} F^b_{jk} F^c_{ki}
- \frac{ N_d g_w^2 D6}{15} (D_i F_{ij})^a (D_k F_{kj})^a \, .
\label{puregauge}
\end{equation}

Fermionic contributions to other independent gauge invariant dimension 6
operators (such as $(A_0^2)^3$) vanish.

Exactly the same terms arise whether the massive quark is top type or
bottom type.  If both types are massive (in one generation) then each
occurrence of $g_t^n$ becomes $g_t^n + g_b^n$ in the above, and exactly one
mixed term appears at dimension 6,
\begin{equation}
- \frac{4 N_c g_t^2 g_b^2 D6}{3}
\Phi^{\dag} D_j i \tau_2 \Phi^* (D_j i \tau_2 \Phi^*)^{\dag} \Phi \, .
\end{equation}

Fermions also induce pure QCD operators and operators containing both weak and
strong fields; these are unimportant here and are listed in Appendix B.

We should comment that, except when they coincidentally vanish, the
fermionic contributions to dimension six operators tend to be much larger
than the contributions from the nonzero Matsubara frequencies of boson fields,
some of which have been worked out in \cite{Patkos}.  This is partly because
the top quark is heavier than any of the bosonic degrees of freedom and
partly because its lowest nonzero Matsubara frequency is $\pi T$, rather
than $2\pi T$.  Because of this the bosonic
equivalent of the coefficient $D6$ is smaller
by a factor of 7.  Hence, we anticipate that if the expansion in high
dimension operators is well behaved for fermions then it will also be
well behaved for bosons.

We have also computed the fermionic contributions to masses, couplings, and
wave function renormalizations in a slight modification of the proper time
technique of DPSSG.  The procedure is outlined in Appendix A.  The
results turn out to be identical to those in $\overline{\rm MS}$ except
that $D4$ is modified to
\begin{equation}
D4_{\rm proper \: time} = \frac{ \gamma_E - 2 \ln \pi + \ln(\mu^2 / T^2)}
{16 \pi^2} \, ,
\label{D4propertime}
\end{equation}
where the definition of the scale $\mu$ is explained in Appendix A.
This expression relates the $\overline{\rm MS}$ and proper time renormalization
points.  The appendix also presents the calculation of the vacuum effective
potential and the physical Higgs mass in the proper time scheme.

\section{The Sphaleron}

We want to apply these results to find the free energy of a nontrivial
field configuration which solves the classical equations of motion.
Klinkhammer and Manton have shown \cite{Klink} that the classical equations
of motion of the $SU(2)$ Higgs system can be solved by an {\it Ansatz}
of form
\begin{equation}
\Phi = \frac{\nu}{\sqrt{2}} \frac{h(r)}{r} \left[ \begin{array}{c} x+iy \\ z
\end{array} \right] \, , \qquad
A_i^j = \frac{2 f(r)}{r^2} r_k \epsilon_{ijk} \, .
\label{Sphaleron}
\end{equation}
The lower index on $A$ is the Lorentz index and the upper index is
the group index, and $h$ and $f$ are functions of $r$ alone which are
to be determined by minimizing the configuration's energy, subject to
the boundary conditions $f(0)=h(0)=0$ and $f(\infty)=h(\infty)=1$.
(Here and throughout $\nu$ is the Higgs VEV,
$\nu^2(T) = 2 \Phi^{\dag} \Phi (T) $.)
This solution is called the Sphaleron, and when $E_{Sph} >> T$, the
formalism of Langer tells us that most
baryon number violating events should occur because of phase space paths
which pass close to this configuration.  Arnold and McLerran have applied
this idea to estimate the rate of baryon number violation in the broken
electroweak phase to be \cite{McLerran}
\begin{equation}
\frac{dN_B}{N_B dt} = -13 N_F T \left( \frac{\alpha_w}{4\pi} \right)^4
\frac{\omega_-}{2 m_W}  \left( \frac{4 \pi \nu}{g_w T} \right)^7
{\cal N}_{\rm tr} {\cal NV}_{\rm rot} \kappa \exp(-E_{Sph}/T)
\label{eraserate}
\end{equation}
with $N_F=3$ the number of generations.
The prefactors arise from the zero and unstable modes of the Sphaleron,
and are evaluated in \cite{CarsonI,Carson}.  At 1 loop $\kappa$ is
a product of fluctuation determinants around the configuration,
\begin{equation}
\kappa = \frac{ {\rm Det} H}{ {\rm Det} H_0} \left( \frac{ {\rm Det} K_0}
{ {\rm Det} K'} \right)^{\frac{1}{2}} \, .
\end{equation}
$H$ is Eq. (\ref{Hamiltonian}) in the Sphaleron background and $H_0$
is Eq. (\ref{Hamiltonian}) in the naive vacuum.  $K_0$ is the bosonic
fluctuation determinant in vacuum, and $K'$ is the bosonic determinant in
the Sphaleron background, but with the zero and unstable modes removed.

We should comment that the division between $\ln \kappa$ and $-E_{sph}/T$ is
somewhat arbitrary.  In particular it is renormalization point dependent.
We see this explicitly from our calculation of the contributions from
fermions to dimension 4 operators (Eq. (\ref{dim4contrib})), which are
part of $-\ln \kappa$, but which have
with coefficients $\propto D4$ which explicitly
depend on $\mu$.  The calculation of $E_{Sph}$ correspondingly depends
on coupling constants $g_w$, $\lambda$ which depend on $\mu$.

To compute $E_{Sph} - T \ln \kappa$ we need to evaluate the tree level
Lagrangian terms $\Phi^{\dag} \Phi$, $(\Phi^{\dag} \Phi)^2$, and
$F^{a}_{ij} F^{a}_{ij}$, and the
operators found in Section II, in the Sphaleron
background, Eq. (\ref{Sphaleron}).  First note that $A_0 = 0$, so all
terms including $A_0$ vanish.  Effective potential terms give
\begin{eqnarray}
\Phi^{\dag} \Phi & = & \frac{\nu^2}{2} h^2,
\\
( \Phi^{\dag} \Phi )^2 & = & \frac{ \nu^4}{4} h^4,
\\
( \Phi^{\dag} \Phi)^3 & = & \frac{ \nu^6}{8} h^6 \ldots \: .
\end{eqnarray}
We will also need the term $ (\Phi^{\dag} \Phi)^{3/2} = \nu^3 h^3 /
(2 \sqrt{2})$ which arises from bosonic fluctuations, and we should add
a $\Phi^{\dag} \Phi$ independent constant to make the global minimum
of $V(\nu)$ zero, to subtract out the energy density in the absence of
a Sphaleron.

The dimension 4 derivative terms are \cite{Klink}
\begin{eqnarray}
F^a_{ij} F^a_{ij} & = & 16 \frac{f'^2}{r^2} + 32 \frac{f^2 (1-f)^2}{r^4} \, ,
 \\
(D_i \Phi)^{\dag} D_i \Phi & = & \frac{\nu^2}{2} \left( 2 \frac{h^2 (1-f)^2}
                  {r^2} + h'^2 \right) \, .
\end{eqnarray}

We calculate that the dimension 6 derivative terms are
\begin{eqnarray}
f_{abc} F^a_{ij} F^b_{jk} F^c_{ki} & = & 96 \frac{f'^2 f (1-f)}{r^4} \, ,
\label{firstone} \\
(D_i F_{ij})^a (D_k F_{kj})^a & = & 8 \frac{f''^2}{r^2} - 32 \frac{f''
              f(1-f)}{r^4} + 32 \frac{f^2 (1-f)^2}{r^6} \, , \\
\Phi^{\dag} \Phi F^a_{ij} F^a_{ij} & = & \frac{\nu^2 h^2}{2} \left(
            16 \frac{ f'^2}{r^2} + 32 \frac{f^2 (1-f)^2}{r^4} \right) \, , \\
(D^2 \Phi)^{\dag} D^2 \Phi & = & \frac{\nu^2}{2} \left( h'' + 2 \frac{h'}{r}
	   -2 \frac{h(1-f)^2}{r^2} \right)^2  \, , \\
\Phi^{\dag} \Phi (D_i \Phi)^{\dag} D_i \Phi &=& \frac{\nu^4}{4} \left(
           2 \frac{h^2 (1-f)^2}{r^2} + h'^2 \right) \, , \\
\partial_i (\Phi^{\dag} \Phi) \partial_i (\Phi^{\dag} \Phi) & = &
           \nu^4 h^2 h'^2 \, , \\
\Phi^{\dag} D_i \Phi (D_i \Phi)^{\dag} \Phi & = & \frac{\nu^4}{4} \bigg(
            \frac{h^4 (1-f)^2}{r^4} (x^2 + y^2) + h^2 h'^2 \bigg) \, .
\label{asphericalone}
\end{eqnarray}
Because $f \propto r^2$ and $h \propto r$ at small $r$,
all of these terms are nonsingular at $r = 0$.
The last term, and only the last term, is not spherically symmetric.
The departure from spherical symmetry arises because the Dirac equation
in the presence of
the Sphaleron is not spherically symmetric when only one flavor of quark
has a mass \cite{DPSSG}.  DPSSG evaded this problem by giving both flavors
equal masses, which restores the symmetry.  As we have seen, giving both
quark flavors masses introduces a new dimension 6 operator, whose
free energy density in the Sphaleron background is
\begin{equation}
\Phi^{\dag} D_i i \tau_2 \Phi^* (D_i i \tau_2 \Phi^*)^{\dag} \Phi  =
        \frac{\nu^4}{4} \frac{h^4(1-f)^2}{r^4} (r^2 + z^2) \, .
\label{extraterm}
\end{equation}
The coefficients of this term and (\ref{asphericalone}) are such that,
when $g_t^2 = g_b^2$, the $z^2$ combines with the $x^2 + y^2$ to restore
the spherical symmetry of the terms.

The approximation of DPSSG differs from the results with only the top quark
massive by the contribution of Eq.(\ref{extraterm}) and is accurate only
when this term is small.  This is the case only when derivative dimension
6 operators give very small contributions, which is precisely the case where
dimensional reduction is accurate.

To compute the Sphaleron energy it is convenient to follow \cite{Klink} and
introduce a dimensionless radial distance $\xi = g_w \nu r$.  The contribution
to the Sphaleron energy from the effective potential is then
\begin{equation}
\frac{4 \pi \nu}{g_w} \int_0^{\infty} d \xi \xi^2
\frac{V(h \nu)}{g_w^2 \nu^4} \, .
\label{Vcont}
\end{equation}

The contribution from kinetic energy terms, including the fermions'
contribution, is
\begin{equation}
\frac{4 \pi \nu}{g_w} \int_0^{\infty} \! \!  d\xi \:
\left( 1 + \frac{ N_d g_w^2 D4}{3} \right) \left( 4 f'^2
+ 8 \frac{f^2(1-f)^2}{\xi^2} \right)
+ \left( 1 + N_c g_t^2 D4 \right) \left( h^2 (1-f)^2
+ \frac{ \xi^2 h'^2}{2} \right)
\label{derivcont}
\end{equation}
where derivatives are with respect to $\xi$.

The contribution from Eq. (\ref{puregauge}) is
\begin{eqnarray}
\frac{4 \pi \nu}{g_w} \frac{7 \zeta(3) g_w^4 \nu^2}{128 \pi^4 T^2}
\int_0^{\infty} d \xi \xi^2 \left(
- \frac{N_d}{180} \left[ 96 \frac{f'^2 f(1-f)}{\xi^4} \right] \right.
\qquad \qquad \qquad \qquad \qquad  \nonumber \\
\left. -  \frac{N_d}{15} \left[ 8\frac{f''^2}{\xi^2}
- 32 \frac{f'' f (1-f)}{\xi^4}
+ 32 \frac{f^2 (1-f)^2}{\xi^6} \right] \right) \, .
\label{puregaugecont}
\end{eqnarray}

We can get the other dimension 6 operators from Eqs.
(\ref{Higgsderivs} - \ref{puregauge})
and Eqs. (\ref{firstone} - \ref{asphericalone}) by always replacing
$D6$ with $7 \zeta(3) g_w^4 \nu^2/ 128 \pi^4 T^2$, removing $g_w$'s,
replacing $g_t$ with $g_t/g_w$ and $r$ with $\xi$, and integrating
over $ (4 \pi \nu /  g_w) \int \xi^2 d\xi$.  For $\nu \sim T$,
$D6$ is very small and dimension 6 operators have the parametric appearance
of two loop effects.

{}From the discussion after the introduction in \cite{McLerran} we see that
we want to find the configuration of form (\ref{Sphaleron}) with
minimum free energy, that is the configuration which maximizes
$(\Pi {\rm zero \; mode \: cont.}) \kappa \exp(-E_{Sph}/T)$.  This is a
formidable task, since we cannot compute all of these terms analytically; in
particular the zero mode contributions and part of the zero Matsubara
frequency bosonic contribution to $\kappa$ are beyond our analytic abilities.
Fortunately, the Sphaleron is a saddlepoint configuration, and we should get
almost exactly the right free energy if we include the dominant effects
in the computation of the field configuration $f(\xi)$ and $h(\xi)$, and then
compute the other corrections holding $f$ and $h$ fixed.  This is because,
as a saddlepoint, the free energy of the Sphaleron only changes
quadratically with small changes to $h$ and $f$.

To illustrate this point consider the Sphaleron configuration
computed from a tree level
Lagrangian $ F^{a}_{ij} F^{a}_{ij}/4$$
 + \lambda (\Phi^{\dag} \Phi - \nu^2 /2 )^2$ and
suppose that we are only interested in the correction fermions induce in the
gauge fields, $N_d g_w^2 D4 F^{a}_{ij} F^{a}_{ij}/12$.
We will choose $\mu = m_W(T=0)$
in the proper time scheme and $T = 100$ GeV, so $D4 \simeq - 0.0135$.
Solving for the Sphaleron using the tree action we find
$E_{Sph} = 35.0975 \nu$,
and estimating the fermionic correction as
\begin{equation}
\frac{4 \pi \nu}{g_w} \int_{0}^{\infty} d \xi \left( \frac{N_d g_w^2 D4}{3}
\right) \left( 4 f'^2 + 8 \frac{ f^2 (1-f)^2}{\xi^2} \right)
\simeq -0.4433 \nu
\end{equation}
we get a total free energy of $34.6542 \nu$.  When we solve for the Sphaleron
configuration, including the fermionic contribution as well as the tree
terms, we find the Sphaleron energy, which now includes the fermionic
correction, is $34.6520 \nu$.  The correction from this fermionic
contribution is about $1\%$ of $E_{Sph}$, and the error in estimating it
at fixed $f$ and $h$ is about $0.01 \%$, which is quadratic, as expected.

Obviously, though, this will not do when a correction is actually
substantial and the modification of the free energy is comparable to
$E_{Sph}$.  This is potentially the case for the fermionic correction to
the $(\Phi^{\dag} \Phi)^2$ term in the effective potential.  For the
above parameters, the coefficient of the correction is $N_c g_t^4 D4
= -0.040$, which is to be compared with $\lambda(\mu) = 0.050$.
Because of an unfortunate choice for $\mu$, the ``correction'' for
top quarks is almost as large as the tree level term itself, and we
certainly cannot trust a result in which $f$ and $h$ are computed with
the tree term and used to compute the quark fluctuation determinant.
The only consistent,
renormalization point independent thing to do is to include those
fermionic contributions which correct operators appearing in the tree
level action in the calculation of the Sphaleron configuration,
that is to use the $\mu$ independent quantity $\lambda(\mu) + N_c g_t^4 D4$
in the calculation of $f$, $h$, and $E_{Sph}$.  This is particularly
important for the scalar self-coupling, because as mentioned earlier it
is unprotected from large radiative corrections and in fact the top
quark contribution here is substantial.

To work in terms of physical quantities, we first find a relation
between the couplings at the scale $\mu$ and vacuum, physical masses.  That
is, we find $\lambda(\mu)$ in terms of $m_H(T=0)$.  We perform this
calculation, including all one loop, fermionic contributions, in
Appendix A.  We ignore the bosonic corrections because we are most
interested in understanding the fermionic radiative corrections in this
paper, and because the bosonic corrections are smaller by a factor
of $9 m_W^4/12 m_t^4 \simeq 1/30$.  From the Appendix A results we find
$\lambda ( \mu ) = (m_H^2/2 \nu_o^2)
- (N_c g_t^4/16 \pi^2) (\ln (\mu^2/m_t^2) - a)$,
where $a=0$ in $\overline{\rm MS}$ and $a = \gamma_E$ in proper time
regulation.
The quantity we should use in calculating the Sphaleron
configuration at a temperature $T$ is
\begin{equation}
\lambda_T \equiv \lambda(\mu) + N_c g_t^4 D4 \, ,
\label{shoulduse}
\end{equation}
which has no $\mu$
or renormalization scheme dependence, as we can note by inspecting
Eq. (\ref{D4def}) or Eq. (\ref{D4propertime})\footnote{The alert
reader may notice that $\lambda(\mu)$ also contains a term
$(m_H^2/2\nu_o^2)( N_c g_t^2/16 \pi^2)( \ln (\mu^2 / m_t^2) - a)$, which
apparently spoils the cancellation of the $\mu$ dependence discussed here.
This is true, and it gets another correction because $\nu_o$ is $\mu$
dependent;
but the mass squared parameter $m_0^2$ of the effective
potential contains a similar $\mu$ dependence, so the effect of the correction
is just to shift the location of the minimum of $V$ by a proportional
amount, without changing its height.
The Higgs field wave function has the same dependence, so this has no
influence on the Sphaleron energy.}.

A particularly convenient choice for
$\mu$ is
\begin{equation}
\mu = \exp(- \gamma_E) \pi T, \; \overline{\rm MS} \quad {\rm or}
\quad
\mu  =  \exp(- \gamma_E / 2) \pi T, \; {\rm proper \; time}
\label{muchoice}
\end{equation}
because in this case $\lambda_T = \lambda(\mu)$ and the Higgs field and
gauge field wave functions take their tree values; but there is no need
to choose this scale, as long as we compute the Sphaleron configuration
and energy using couplings and wave functions corrected by the fermion
contributions as in equations (\ref{derivcont},\ref{shoulduse}).
This is precisely the
prescription of the dimensional reduction program.  We can test its
reliability by seeing how large the remaining corrections, those coming
from dimension 6 operators, are.

Of course, it is also necessary to
include corrections from zero Matsubara frequency bosons, which are
expected to be considerable and cannot be evaluated with a derivative
expansion, as we have already discussed.  The largest part of this
correction comes from the cubic effective potential terms.  The residual
correction when this is completely removed is $\ln \kappa \simeq 1.5$
\cite{Junker,Junker2} which is small enough that we can trust the
computation in terms of fixed $f$ and $h$ as discussed above; but
the effective potential contribution is large (at $T \simeq T_c$ it
changes the very nature of the phase transition and should be considered
an order 1 correction to $E_{Sph}$) and should be included in the
evaluation of the Sphaleron configuration, as advocated in \cite{Shap}.
But before we can compute
the Sphaleron energy including these corrections we must discuss the
nature of the phase transition to find out at what temperature the
baryon erasure begins.

\section{Phase transition, Sphaleron energy}

Let us briefly review the electroweak phase
transition.  At one loop the zero Matsubara frequency bosonic excitations
generate negative cubic terms in the effective potential, which becomes
($\Theta_W = 0$)
\cite{Arnold}
\begin{eqnarray}
V( \nu ) = - \frac{g_* \pi^2 T^4}{90} +
\left( - \frac{m_0^2}{2} + \frac{(4g_t^2+3 g_w^2 + 8 \lambda)T^2}{32}
    \right) \nu^2
 - \frac{g^3 T}{16 \pi} \nu^3
   - \frac{ g_w^3 T}{4 \pi}
   \left( \frac{11 T^2}{6} + \frac{\nu^2}{4} \right)^{\frac{3}{2}}
\nonumber \\
   - \frac{3 m_2^3 + m_1^3}{12 \pi} + \lambda_T \frac{\nu^4}{4}
+{\rm dimension \; six} \, , \quad
\end{eqnarray}
where $m_1^2 = \lambda_T \nu^2/2 + V''(\nu = 0)$,
$m_1^2 =  \lambda_T \nu^2 + m_2^2$, $g_* = 106.75$ is the number
of radiative degrees of freedom, and ``dimension six'' means the higher terms
found in Eq. (\ref{hiTexpan}).
At high temperature $V$ has only one minimum
at $\nu = 0$, but as temperature drops the negative cubic terms generate
a second ``asymmetric'' minimum.  At some temperature the second minimum
becomes more thermodynamically favorable, and bubbles of the asymmetric
phase begin to nucleate and grow shortly thereafter.  The temperature at
which the nucleations become common can be computed by standard techniques
\cite{Dine}.  The expanding bubbles liberate latent heat, so that the
temperature after the transition is somewhat higher than the nucleation
temperature.  The temperature the plasma reheats to due to this latent heat
is determined by the condition that the broken phase energy density
$E = V - T \partial V / \partial T$
must equal the symmetric phase energy density at the temperature where
the nucleations occurred.
It is at this reheat temperature, immediately after the phase
transition, that quasi-equilibrium erasure of any baryon number excess
generated at the phase transition begins.  It continues for all times
thereafter, but as Eq. (\ref{eraserate}) shows,
the rate depends strongly on the
Sphaleron energy, which changes rapidly with $T$, as $\nu$ moves towards
its zero temperature value.  Almost all the baryon erasure takes place within
a fraction of a Hubble time, so the Sphaleron
rate is only relevant at temperatures quite close to the reheat temperature.

We have computed the reheat temperature and the Sphaleron energy at the
reheat temperature for a number of physical Higgs masses, using the 1
loop relations between Higgs mass and $\lambda$ presented in Appendix A and
including the negative cubic effective potential terms from zero Matsubara
frequency bosonic modes.
(We always use $g_w = 0.65$ and take the Weinberg angle to be zero.  In fact,
to account for the quark contribution to the gauge
field wave function, the value of $g_w$ we should use is
$g_w(\mu) - N_d g_w^3 D4/6$, which is $T$ dependent.  The $T$ dependence
is very mild, behaving as $g_w * (N_d g_w^2 / 48 \pi^2 \simeq 0.01) \ln T$.
Since this dependence is so weak, and
since we are not including the influence of bosons' nonzero Matsubara modes,
which will contribute an opposite and slightly larger temperature dependence
\cite{FKRS}, we will not worry about it.  However we will take full
consideration of the correction to $\lambda$, which as discussed is not at
all weak.)
The results, together with the contributions from
dimension 6 operators, are presented in table 1.  The $\nu^6$ contribution
to the effective potential was computed by inclusion in the calculation of
the Sphaleron configuration, and the others were performed perturbatively.
It is clear from the table that, as expected,
dimension 6 operators make only a tiny contribution to the Sphaleron energy.
Hence we can conclude that near the phase transition temperature the
expansion is very well behaved and the use of the dimensionally reduced
theory is well justified.  The largest dimension 6 correction comes from
the effective potential term, owing to the high power of $g_t$; this and
only this term may not be completely negligible, increasing
the Sphaleron energy and the strength of the phase transition
by a few percent for light Higgs.
Generally the terms become progressively less important as they contain more
derivatives.  In particular the very small contribution from
$(D^2 \Phi)^{\dag}) D^2 \Phi$ gives confidence that the expansion of
Eq. (\ref{derivexpan}) in derivatives is justified.

We have not continued the table down below 30 GeV partly because this
range is experimentally excluded, partly because questions of vacuum
stability become increasingly hard to avoid in this range, and partly
because the results barely differ from those at $m_H = 30$ GeV.
This is because the one
loop relations for the Higgs mass give nonzero $\lambda_T$ even as the
physical Higgs mass $m_H \rightarrow 0$.

The next step is to use these results to determine the baryon number
depletion.  From Eq.(\ref{eraserate}) we find that baryon number is depleted
by a factor of
\begin{eqnarray}
\exp \left(
  \int_{t_{\rm reheat}} 13N_F T \left( \frac{\alpha_w}{4 \pi} \right)^4
\frac{\omega_-}{2 m_W} {\cal N}_{tr} {\cal NV}_{rot}
\left( \frac{4 \pi \nu}{g_w T} \right)^7 \kappa \exp \frac{-E_{Sph}}{T} dt
\right)
\label{erasure}
\end{eqnarray}
The elapsed time is related to the change in temperature by $dt = dT/(HT)$,
with $H$ the Hubble constant, which is approximately $H = T(8 \pi^3 g_*/90)
^{1/2} (T/m_{pl})$.  To perform the integral
we must numerically repeat the evaluation of the 
Sphaleron energy at many values of $T$ close to the reheat temperature.
Most of the coefficients in Eq. (\ref{erasure}) are given in 
\cite{CarsonI}.  We use the values found there for $\omega_-$ and
the ${\cal N}$, even though they were computed for slightly different $f$
and $h$, because the product of 
these factors turns out to be very insensitive
to changes in the configuration \cite{CarsonI}.
The other factor we require is the difference between the 
Sphaleron energy which we have calculated and the (1 loop) value of 
$E_{Sph} - T \ln \kappa$.  This is due to derivative corrections from 
zero Matsubara frequencies and can be approximated from the results
of Baacke and Junker \cite{Junker}, who find that, when the full tadpole
is removed (ie. all effective potential contributions are subtracted out),
$\ln \kappa$ is about 1.5, independent of $\nu$ and quite weakly dependent
on $\lambda_T$.  (Again, this quantity was computed for different $f$, $h$ in
that paper, but again it proved to be quite insensitive to configuration,
so we will use this value.  This introduces an uncertainty in our final value
for the erasure rate of perhaps $\pm 1$ in the exponential.)  
We can then perform the integral in
Eq. (\ref{erasure}) numerically.  We find that almost all of the 
erasure occurs in a range of $T$ less than $0.5 \% $ from the reheat
temperature, essentially because $\nu$ changes very rapidly with temperature
immediately after the transition.  This narrows the available time for the
Sphaleron erasure, but even for 
$m_H =$30 GeV we find that the baryon number is depleted by a factor of 
about $\exp(11.3)$.  If the conjecture about the effect of Landau damping
on the negative frequency mode in \cite{McLerran} is correct then the
suppression is smaller by about 2 in the exponent.
The results for several Higgs masses are presented
in the table; in all cases the erasure is very considerable.  Since the
most optimistic estimates of baryogenesis in the minimal standard model
can barely account for the current abundance \cite{Farrar}, this
apparently rules out electroweak physics as the source of baryogenesis
in the minimal standard model.

\section{Conclusion}

It appears that a perturbative treatment of fermions is very well justified
near the phase transition temperature,
and that dimensional reduction should be accurate, though the correction 
{}from the dimension 6 contribution to the effective potential may not be
completely negligible.  

How should we understand the results of DPSSG in
light of this conclusion?  

First recall what we have done here.  We find
that at a general $\mu$ the fermions will induce nonzero corrections to
couplings and wave functions, and in particular the correction to $\lambda$
is potentially large.  Following the idea of the dimensional reduction
program we combine these contributions with the tree couplings, resulting
in renormalization point independent couplings $\lambda_T$, which are used
to compute the Sphaleron energy.  The Sphaleron energy then already contains
that part of the fermion fluctuation determinant which is understood as
coupling and wave function corrections; the residual correction, which 
comes from dimension 6 (and higher) operators, is explicitly found to be
very small.  However, had we used the tree couplings at some scale $\mu$,
we would then expect fermions to give a ($\mu$ dependent) nonzero
correction due to the difference $\lambda_T - \lambda(\mu)$.   Only at
one particular, convenient renormalization point (Eq. (\ref{muchoice}))
would this contribution vanish.

Diakonov et. al. use a fixed $\mu$ (in \cite{DPSSG} they use $\mu=m_W$
in the proper time scheme,
and in \cite{DPSSG2} they use $\mu \simeq m_t \exp(\gamma_E/2)$) and
compute the Sphaleron configuration from the tree Lagrangian (corrected 
however by the thermal contributions to the $\Phi^{\dag} \Phi$ term).  We
then expect from our work that they should find a correction arising
from the difference $\lambda_T - \lambda(\mu)$ equal to
\begin{equation}
\frac{4 \pi \nu}{g_w} \frac{\lambda_T - \lambda(\mu)}{g_w^2}
\int_{0}^{\infty} d\xi \xi^2 \frac{h^4 - 1}{4} \, ,
\end{equation}
which will depend on the top mass as $g_t^4$, exactly as they find.  This is
not a contradiction of the dimensional reduction scheme, which predicts
that, because of their use of $\lambda(\mu)$ with their choice of $\mu$, 
they should find such a term.  We should note that, when this term is
large (as it is for the choice of $\mu$ made in \cite{DPSSG}), one
should not trust the functions $h$, $f$ computed from the tree
Lagrangian but should include this radiative correction in their computation,
as we have done here.  In other words, the numerical work of 
Diakonov et. al. is probably accurate, but because of the way they
have done the problem they will not necessarily produce the free energy
of the configuration which actually limits the baryon erasure rate.

We should also note that in \cite{DPSSG} Diakonov et. al. 
use a tree, rather than 1 loop,
relation between $m_H$ and $\lambda(\mu)$, and a tree, rather than 1 loop,
value of $\lambda$ in calculating 
the phase transition temperature.  (In \cite{DPSSG2} the 
relation between $m_H$ and $\lambda(\mu)$ is computed at one loop, but the
transition temperature is still found using $\lambda(\mu)$ rather than 
$\lambda_T$.)  This is inconsistent with a one loop analysis of the
Sphaleron rate and may explain the difference in our
results for the dissipation of baryon number.  

What is the overall effect of fermions?  For small Higgs mass, 
$\lambda_T > \lambda_{\rm tree}$,
and as $\nu(T_{\rm reheat})$, and hence $E_{Sph}$, fall with increasing
$\lambda_T$, we find more baryon number dissipation than we would if we
ignored fermions altogether and used $\lambda_{\rm tree}$.  This is the
reason that, even for very small Higgs mass, we still find substantial
baryon number erasure.  We should emphasize once more that to find this
result it was important to apply one loop corrections systematically, in
the effective potential, the phase transition temperature, and the Sphaleron
energy, but that the effect is basically perturbative, and the high 
temperature expansion accounts for it successfully.

Do the results of the last section preclude electroweak baryogenesis?
They make it unlikely that baryogenesis can be viable in the minimal
standard model.  However we should note that we have only used the 1 loop
effective potential, and while extending our results to the two loop 
potential, which is known, gives essentially the same conclusions, it is
not clear that the perturbative treatment of the effective potential is
reliable; the phase transition may be stronger than perturbation theory
suggests.  Also, we have said nothing about extensions to the standard model.
For instance, in the two doublet model, the phase transition can be much
stronger without contradicting experimental bounds on the Higgs mass
\cite{someone}, and the Sphaleron bound only narrows the parameter space.

\centerline{Acknowledgements}

I am very grateful to Misha Shaposhnikov for useful conversations,
correspondence, and encouragement, and to Mikko Laine, for comparing
unpublished calculations of the fermionic contributions to dimension 4
operators with me.  I also acknowledge funding from the NSF.

\section{Appendix A:  Proper Time Calculations}

In this appendix we will show how to compute the finite temperature
fluctuation determinant using a proper time regulation analogous to that of
DPSSG.  The basic idea is to use the relationship
\begin{equation}
\ln \frac{ {\rm Det} K}{ {\rm Det} K_0} = \lim_{\epsilon \rightarrow0}
- { \rm Tr} \int_{\epsilon}^{\infty} \frac{dt}{t}
(e^{-Kt} - e^{-K_0t}) \, ,
\end{equation}
which holds when $K$ is a quadratic, positive definite operator and $K_0$
is its free, massless approximation.  In our case the operator $\gamma^\mu
\partial_\mu - m$, or Eq. (\ref{Hamiltonian}), is not quadratic or positive
definite.  Fortunately it has the same spectrum as the operator
$-\gamma^\mu \partial_\mu - m$, so we can write the determinant we actually
want in terms of a quadratic, positive definite operator,
\begin{eqnarray}
 -\ln {\rm Det}( \gamma^\mu \partial_\mu - m)
 & = & -\frac{1}{2} \left(
\ln {\rm Det}( \gamma^\mu \partial_\mu - m)
+ \ln {\rm Det} ( -\gamma^\mu \partial_\mu - m) \right)
\nonumber \\
 & = & - \frac{1}{2} {\rm Tr} \ln ( -\gamma^\mu \gamma^\nu \partial_\mu
\partial_\nu - \gamma^\mu (\partial_\mu m) + m^2) \, ,
\end{eqnarray}
which acts on a Euclidean space where the time direction is cylcic with
antiperiodic boundary conditions and period $T$.  It therefore already
includes all thermal effects, which do not have to be added by hand,
as in the treatment of DPSSG.  However, we will have to be careful
when we regulate to see that we are making $T$ independent subtractions.

Following \cite{DPY} we perform the trace by inserting a complete set
of plane-wave states,
\begin{equation}
{\rm Tr} e^{-Kt} = {\rm tr} \int d^4 x T \sum_{p_0} \int \frac{d^3
p}{(2\pi)^3} e^{-i p \cdot x} e^{-Kt} e^{i p \cdot x} \, ,
\end{equation}
where tr is over Dirac indicies.  The factor $\exp(i p \cdot
x)$ can be brought through the operator to cancel $\exp(-ip \cdot x)$,
but in doing this all derivative operators in $K$ are shifted,
$\partial_\mu \rightarrow \partial_\mu + i p_\mu$.  The zero
temperature limit is found by replacing the sum on $p_0$ with the
integral $\int dp_0/(2 \pi)$.

We illustrate the renormalization procedure by computing the effective
potential in this regulation.  In this case $K = - \partial^2 + m^2$
because the mass is space independent.  After the shift, the
derivatives have nothing on which to act and do not contribute; we can
drop them.  The problem is then to compute
\begin{equation}
2 \int_{\epsilon}^{\infty} \frac{dt}{t} \int d^3x
 \sum_{p_0} \int \frac{d^3 p}{(2 \pi)^3}
e^{- p^2 t} e^{-p_0^2 t} (e^{-m^2 t} - 1) \, ,
\end{equation}
where we have performed the Dirac trace and removed the trivial integral
over $dx_0$ so that the
result will be the free energy density.

Next we expand $\exp(-m^2 t)$ in powers of $t$.  The first term is
cancelled by the $-1$.  The second gives
\begin{equation}
-2\int m^2 d^3 x   \int_{\epsilon}^{\infty} dt T \sum_{p_0} e^{- p_0^2 t} \int
\frac{d^3 p}{(2 \pi)^3} e^{-p^2 t} \, .
\end{equation}

This expression is small $t$ divergent, cut off by $\epsilon$.  To
render the theory cutoff independent we should add and subtract a
temperature independent expression with the same small $t$ behavior
and absorb the one we added with a counterterm in the tree level mass.
The correct expression to subtract is
\begin{equation}
-2 \int m^2 d^3 x  \int_{\epsilon}^{\infty} dt \int \frac{dp_0}{2 \pi}
e^{-p_o^2 t} \int \frac{d^3 p}{(2\pi)^3} e^{-p^2 t} \, .
\end{equation}

There is now no obstacle to performing the integral over $t$, or to
setting $\epsilon$ equal to zero.  The result is
\begin{equation}
-2\int m^2 d^3 x \int \frac{d^3 p}{(2 \pi)^3} \left( T \sum_{p_0}
- \int \frac{dp_0}{2 \pi} \right) \frac{1}{p^2 + p_0^2} \, .
\end{equation}
Both the sum and the integral over $p_0$ are straightforward, giving
\begin{equation}
\sum_{p_0} \frac{1}{p_0^2 + p^2} = \frac{ {\rm tanh}
\frac{p} {2T}}{2p} \, , \quad \int \frac{dp_0}{2\pi} \frac{1}{p_0^2 + p^2}
= \frac{1}{2p} ', ;
\end{equation}
combining them and performing the integral over angles, we get
\begin{equation}
-2 \int d^3x m^2 \frac{1}{2 \pi^2} \int \frac{-1}{\exp(p/T) + 1} p dp
= \int d^3 x \frac{m^2 T^2}{12} \, ,
\end{equation}
which is the well known expression for the thermal contribution to the
Higgs mass squared.

In their discussion of the renormalization of the theory DPSSG
advocate cutting off the proper time integral of the counterterm at
some finite upper bound, call it $\mu^{-2}$ (in their case, $m_W^{-2}$).
Doing so changes the result of the above calculation to
\begin{equation}
\frac{m^2 T^2}{12} - \frac{\mu^2 m^2}{8 \pi^2} \, ,
\end{equation}
which means there is a discrepancy between the tree level mass squared
parameter and the renormalized, vacuum mass squared parameter.  There
is nothing wrong with this in principle as long as we remember it is
there; we should use the renormalized mass squared when performing
calculations such as the Sphaleron configuration and remember that we
have already included part of the fermion contribution by doing so; we
will need to subtract it off, along with the thermal mass squared.  It
is much easier and more straightforward, however, to follow our
procedure and absorb the entire vacuum correction in a counterterm.

Continuing to expand $\exp(-m^2 t)$, the next term is
\begin{equation}
+\int m^4 d^3x \int_{\epsilon}^{\infty} t dt \sum_{p_0}
 e^{-p_0^2 t} \int \frac{d^3 p}{(2\pi)^3} e^{-p^2 t} \, ,
\end{equation}
which is logarithmically divergent at small $t$.  Because the lowest
Matsubara frequency is nonzero, it is cut off exponentially at large
$t$ and there are no infrared problems in its evaluation.  Again we
have an available Lagrangian parameter in which to absorb the
divergence, and we should add and subtract a $T$ independent
expression with the same ultraviolet behavior, and absorb the one with
the same sign into the Higgs self-coupling parameter.  The
corresponding vacuum integral is both ultraviolet and infrared
divergent, and to prevent infrared divergences we
must introduce a renormalization scale into the problem.  Following
DPSSG, we subtract
\begin{equation}
\int m^4d^3 x \int_{\epsilon}^{\mu^{-2}} t dt \int \frac{dp_0}{2 \pi}
e^{-p_0^2 t} \int \frac{d^3 p}{ (2\pi)^3} e^{-p^2 t} \, ,
\end{equation}
giving
\begin{equation}
\int m^4 d^3x \left( \int_{\epsilon}^{\infty} dt T \sum_{p_0} -
\int_{\epsilon}^{\mu^{-2}} dt \int \frac{dp_0}{2\pi} \right)
 t e^{-p_o^2t} \int\frac{ d^3 p}{ (2\pi)^3} e^{-p^2 t} \, .
\end{equation}
The integral over $d^3p$ gives $t^{-3/2}/( 8 \pi^{3/2})$.  The
counterterm can then be evaluated directly and gives $+ \ln( \mu^2
\epsilon ) / (16 \pi^2)$.  The remaining integral,
\begin{equation}
 \frac{T}{8 \pi^{ \frac{3}{2}}}\sum_{p_0} \int_{\epsilon}^{\infty}
 t^{- \frac{1}{2}} dt e^{-p_0^2 t} \, ,
\label{toughint}
\end{equation}
is more delicate.  For small terms in the sum the integral over $t$ is
approximately $\sqrt{\pi} / p_0$, so the early terms in the sum are
\begin{equation}
\frac{T}{4 \pi} \sum_{l=1,3...} \frac{1}{l \pi T} \, .
\end{equation}
The difference between this sum and a corresponding integral is
concentrated in the first few terms.  The sum is approximately
\begin{equation}
\frac{1}{4 \pi} \left[ \left( \int_{\pi T} \frac{dp_0}{2\pi}
\frac{1}{p_0} \right) + \frac{ \gamma_E + \ln 2}{2 \pi} \right] \, ,
\end{equation}
so smoothing over the sum in Eq. (\ref{toughint}) introduces a
correction of $(\gamma_E + \ln 2)/(8 \pi^2)$, giving
\begin{equation}
\frac{\gamma_E + \ln 2}{8 \pi^2} + \frac{1}{4 \pi^{\frac{3}{2}}} \int
_{\pi T}^{\infty} \frac{dp_0}{2 \pi p_0} \int_{\epsilon}^{\infty} \frac
{dt} {t^{\frac{1}{2}}} e^{-p_0^2 t} \, .
\end{equation}
The integral over $p_0$ can be performed by parts, giving
\begin{equation}
\frac{\gamma_E + \ln 2}{8 \pi^2} + \frac{1}{4 \pi^{ \frac{3}{2}}}
\frac{-1} {2 \sqrt{\pi}} \left( \ln \pi T + \frac{1}{2} \ln \epsilon -
\frac{1}{2} \psi(1/2) \right) \, ,
\end{equation}
with $\psi$ the digamma function, $\psi(1/2) = -\gamma_E - 2 \ln
2$.  The result is then
\begin{equation}
\int m^4 d^3 x \frac{ \gamma_E - 2 \ln \pi + \ln (\mu^2/T^2)}{16\pi^2} \, .
\end{equation}
The fraction in the integral gives the proper time renormalization value
of $D4$.

The terms higher order in $m^2$ are actually easier; the term at order
$m^{2n}$ is given by the integral
\begin{equation}
2 \int_{0}^{\infty} \frac{dt}{t} \int d^3 x \frac{(-1)^n}{n!} t^n m^{2n}
\sum_{p_0} \int \frac{d^3 p}{(2 \pi)^3} e^{-p^2 t} e^{-p_0^2 t}\, .
\end{equation}
Performing the integral over $t$, we get
\begin{equation}
\frac{2 (-1)^n}{n} \int d^3 x m^{2n} \sum_{p_0}
\int_{0}^{\infty} \frac{p^2 dp}{2 \pi^2} \frac{1}{(p^2 + p_0^2)^n} \, .
\end{equation}
The result per unit volume is
\begin{equation}
\frac{ (-1)^n m^{2n}}{n \pi^2} \left( \sum_{p_0} p_0^{3-2n} \right)
\int_{0}^{\infty} \frac{y^2 dy}{(y^2 + 1)^n} \, ,
\end{equation}
from which Eq. (\ref{hiTexpan}) follows immediately by performing the sum
and the integral.  Note that the calculation was completely large $t$ finite
because the $\exp(-p_0^2 t)$ term always decays exponentially fast, since
$p_0^2$ is always at least $(\pi T)^2$.  This is to be contrasted with
the zero temperature \cite{DPY} and zero Matsubara frequency bosonic
\cite{CarsonI} cases and explains why the expansion in operator dimension
is possible at finite temperatures for fermions, while it is known to
have trouble at zero temperature and for the zero Matsubara frequency bosonic
modes.

The same basic techniques can be used for the case with gauge fields
and spatial variation.
A complication which arises when computing other dimension 4 operators
is that, in contributions involving $A_0$, powers of $p_0$
appear in addition to $\exp(-p_0^2 t)$.  The result is that the
argument of the digamma function shifts by 1 per power of $p_0^2$.  This
is why the dimension 4 terms containing $A_0$ are not simply multiples of $D4$.

Next we will compute the fermionic contribution to the vacuum
effective potential in this regulation.  From the above,
the contribution from each color is
\begin{equation}
2 \left[ \left( \int_{\epsilon}^{\infty} \frac{dt}{t}
\int \frac{d^4 p}{(2 \pi)^4} e^{-p^2 t} (e^{-m^2 t} - 1 + m^2 t) \right)
 - \left( \int_{\epsilon}^{\mu^{-2}} \frac{dt}{t} \int \frac{d^4p}{(2\pi)^4}
  e^{-p^2 t} \frac{m^4}{2} t^2 \right) \right] \, .
\end{equation}
The $-1$ is the vacuum energy subtraction, the $m^2 t$ is the mass squared
counterterm, and the last expression is the self-coupling counterterm.
The integrals over $p$ may be performed immediately, giving
\begin{equation}
\frac{1}{8\pi^2} \left(\int_{\epsilon}^{\infty} \frac{dt}{t^3}
 (e^{-m^2 t} - 1 + m^2 t) \right)
- \frac{1}{8 \pi^2} \frac{m^4}{2} \int_{\epsilon}^{\mu^{-2}} \frac{dt}{t} \, .
\end{equation}
After integrating the first expression by parts three times, we get
\begin{equation}
\frac{m^4}{16 \pi^2} \left( \frac{3}{2} + \ln \frac{\mu^2}{m^2} - \gamma_E
\right) \, .
\label{Veffpropertime}
\end{equation}
Recalling that $m^2 = g_t^2 \Phi^{\dag} \Phi$, summing on colors, and adding
this expression to the tree level effective potential
\begin{equation}
-m_0^2 \Phi^{\dag} \Phi + \lambda (\Phi^{\dag} \Phi)^2
\end{equation}
we find that the curvature of the effective potential (the second derivative
with respect to $\nu$) at the minimum ($\nu = \nu_o$) is
\begin{equation}
V'' = 2 \nu_o^2  \left( \lambda + \frac{N_c g_t^4}{16 \pi^2}
( \ln \frac{\mu^2}{m_t^2} - \gamma_E ) \right)  \, ,
\label{Vdubprime}
\end{equation}
and the effective potential parameter $m_0^2$ is
\begin{equation}
 m_0^2 = \frac{V''}{2} + \frac{N_c g_t^4 \nu^2}{16 \pi^2} \, .
\end{equation}
The analogous $\overline{\rm MS}$ result is the same but with the
$- \gamma_E$ removed from Eq. (\ref{Vdubprime}).

$V''$ is not the physical Higgs mass squared.
The physical Higgs mass is the ratio
of potential to kinetic energy of a long wavelength fluctuation about the
minimum, times the wave number of the fluctuation; there is a wave function
correction which requires computing the self-energy at the pole mass.
We have been unable to do the calculation in proper time regulation
(see however \cite{DPSSG2}), but we have performed it in $\overline{\rm MS}$
using an integral from \cite{Bohm}.  The result is that
\begin{equation}
m_H^2 = V'' \left( 1 - \frac{N_c g_t^2}{16 \pi^2}
\bigg( \ln \frac{ \mu^2}{m_t^2} + 2
 - \frac{2 \sqrt{4 m_t^2 - m_H^2}}{m_H} {\rm arctan} \frac{m_H}
	{\sqrt{4 m_t^2 - m_H^2}} \; \bigg) \right) \, .
\label{mHpropertime}
\end{equation}
The proper time value is presumably the same but with a $-\gamma_E$ inserted
after the log.  For $4m_t^2 >> m_H^2$ the expression following the log is
about $2 m_H^2/3(4m_t^2 - m_H^2) \simeq 0$ and we have permitted ourselves
to neglect it when relating the quartic coupling to the physical Higgs mass.

\section{Appendix B:  Other high dimension operators}

We list here those dimension 4 and 6 operators induced by fermions
which contain both weak $SU(2)$ and strong $SU(3)$ fields.
These are probably of no phenomenological consequence and
almost certainly do not influence the strength of the phase
transition.  We include them only for completeness.

We write the time component of the gluon field as $A_{0g}$ and the
field strength tensor as $G_{ij}$.  Denoting the number of families
as $N_F = 3$, the dimension four, mixed terms are
\begin{equation}
\frac{-g_t^2 g_s^2}{8 \pi^2} A_{0g}^2 \Phi^{\dag} \Phi
- \frac{g_w^2 g_s^2 N_F}{16 \pi^2} A_0^2 A_{0g}^2 \, .
\end{equation}

The dimension six mixed terms are
\begin{eqnarray}
2 g_t^2 g_s^2 D6 A_{0g}^2 (D_i \Phi)^{\dag} D_i \Phi
+ g_t^4 g_s^2 D6 A_{0g}^2 (\Phi^{\dag} \Phi)^2
+ \frac{g_s^2 g_w^2 g_t^2 D6}{8} A_{0g}^2 A_0^2 \Phi^{\dag} \Phi
\nonumber \\
+ \frac{g_s^2 g_w^2 N_F D6}{3} A_{0g}^2 (D_i A_0)^a (D_i A_0)^a
+ \frac{g_s^2 g_w^2 N_F D6}{6} A_{0g}^2 F^a_{ij} F^a_{ij}
\nonumber \\
+ \frac{g_s^2 g_w^2 N_F D6}{3} (D_i A_{0g})^a(D_i A_{0g})^a A_0^2
+ \frac{2 g_s^2 g_t^2 D6}{3} \partial_i (A_{0g}^2)
      \partial_i (\Phi^{\dag} \Phi)
\nonumber \\
+ \frac{g_s^2 g_w^2 N_F D6}{3} \partial_i(A_{0g}^2) \partial_i (A_0^2)
- \frac{g_s^2 g_t^2 D6}{3} G^a_{ij} G^a_{ij} \Phi^{\dag} \Phi
+ \frac{ g_w^2 g_s^2 N_F D6}{6} G^a_{ij} G^a_{ij} A_0^2 \, .
\end{eqnarray}

There are also pure glue terms which are identical to the pure gauge
terms listed in the text, but with the substitution $N_d \rightarrow
2 N_f$ where $N_f = 6$ is the number of quark flavors.  The 2 is because
both right and left handed quarks couple to the gluons, wereas the
weak coupling is chiral.

\pagebreak


\begin{table}
\begin{tabular}{ | c | c | c | c | c |} \hline
\multicolumn{2}{|c|}{ $m_H$(GeV)} &  30  &  50  &  60
  \\ \hline
\multicolumn{2}{|c|}{ $T_{\rm reheat}$(GeV) }
& 71.2127 & 86.6495 & 95.6517
\\ \hline
\multicolumn{2}{|c|}{ $E_{Sph}/T$ } & 29.18 & 24.79 & 22.38
 \\ \hline
 & $(\Phi^{\dag} \Phi)^3$ &
 1.6 & 0.60 &
 0.31
 \\ \cline{2-5}
 & $\Phi^{\dag} \Phi (D_i \Phi)^{\dag} D_i \Phi$ &
 $-6.3 \times 10^{-3}$ & $-3.7 \times 10^{-3}$ &
 $-2.7 \times 10^{-3}$
 \\ \cline{2-5}
 & $\partial_i (\Phi^{\dag} \Phi) \partial_i (\Phi^{\dag} \Phi)$ &
 $-3.3 \times 10^{-3}$ & $-1.8 \times 10^{-3}$ &
 $-1.3 \times 10^{-3}$
 \\ \cline{2-5}
 & $(D_i \Phi)^{\dag} \Phi \Phi^{\dag} D_i \Phi$ &
 $-2.1 \times 10^{-3}$ & $-1.2 \times 10^{-3}$ &
 $-8.6 \times 10^{-4}$
 \\ \cline{2-5}
 $E_{Sph}/T$ from & $F^a_{ij} F^a_{ij} \Phi^{\dag} \Phi$ &
 $-9.9 \times 10^{-4}$ & $-6.2 \times 10^{-4}$ &
 $-4.7 \times 10^{-4}$
 \\ \cline{2-5}
 & $(D_i F_{ij})^a (D_k F_{kj})^a$ &
 $-4.6 \times 10^{-4}$ & $-2.9 \times 10^{-4}$ &
 $-2.2 \times 10^{-4}$
 \\ \cline{2-5}
 & $ f_{abc} F_{ij}^a F_{jk}^b F_{ki}^c $ &
 $-8.4 \times 10^{-5}$ & $-5.3 \times 10^{-5}$ &
 $-4.0 \times 10^{-5}$
 \\ \cline{2-5}
 & $(D^2 \Phi)^{\dag} D^2 \Phi$ &
 $-4.1 \times 10^{-6}$ & $-4.2 \times 10^{-6}$ &
 $-4.4 \times 10^{-6}$
 \\ \hline
\multicolumn{2}{ | c | }{ depletion factor } & exp(11.3) & exp(15.0) & 
exp(17.1) \\ \hline
\end{tabular}
\vspace{0.2 in}
\begin{tabular}{ | c | c | c | c | c |} \hline
\multicolumn{2}{|c|}{ $m_H$(GeV)}
 & 70 & 80 & 90 \\ \hline
\multicolumn{2}{|c|}{ $T_{\rm reheat}$(GeV) }
 & 105.1149 & 114.8275 & 124.6283\\ \hline
\multicolumn{2}{|c|}{ $E_{Sph}/T$ }
 & 20.49 & 19.26 & 18.70 \\ \hline
 & $(\Phi^{\dag} \Phi)^3$ &
 0.17 & 0.100
& 0.065  \\ \cline{2-5}
 & $\Phi^{\dag} \Phi (D_i \Phi)^{\dag} D_i \Phi$ &
 $-2.0 \times 10^{-3}$ & $-1.6 \times 10^{-3}$
& $-1.5 \times 10^{-3}$  \\ \cline{2-5}
 & $\partial_i (\Phi^{\dag} \Phi) \partial_i (\Phi^{\dag} \Phi)$ &
 $-8.8 \times 10^{-4}$ & $-6.7 \times 10^{-4}$
& $-5.6 \times 10^{-4}$ \\ \cline{2-5}
 & $(D_i \Phi)^{\dag} \Phi \Phi^{\dag} D_i \Phi$ &
 $-6.3 \times 10^{-4}$ & $-4.9 \times 10^{-4}$
& $-4.3 \times 10^{-4}$ \\ \cline{2-5}
 $E_{Sph}/T$ from & $F^a_{ij} F^a_{ij} \Phi^{\dag} \Phi$ &
 $-3.6 \times 10^{-4}$ & $-3.1 \times 10^{-4}$
& $-2.9 \times 10^{-4}$ \\ \cline{2-5}
 & $(D_i F_{ij})^a (D_k F_{kj})^a$ &
 $-1.7 \times 10^{-4}$ & $-1.5 \times 10^{-4}$
& $-1.4 \times 10^{-4}$ \\ \cline{2-5}
 & $ f_{abc} F_{ij}^a F_{jk}^b F_{ki}^c $ &
 $-3.1 \times 10^{-5}$ & $-2.6 \times 10^{-5}$
& $-2.5 \times 10^{-5}$ \\ \cline{2-5}
 & $(D^2 \Phi)^{\dag} D^2 \Phi$ &
 $-4.8 \times 10^{-6}$
& $-5.5 \times 10^{-6}$ & $-7.0 \times 10^{-6}$ \\ \hline
\multicolumn{2}{ | c | }{ depletion factor} & exp(19.0) & exp(20.3) & 
exp(21.5) \\ \hline
\end{tabular}
\caption{Reheat temperature, Sphaleron energy at the reheat temperature,
and contribution of dimension
6 operators to Sphaleron energy for a number of physical Higgs masses}
\end{table}

\begin{figure}
\vspace{2.5in}
\caption{Feynman diagrams corresponding to two insertions of $H_I$.}
\label{one}
\end{figure}


\begin{thebibliography}{99}
\bibitem{Linde} D. Kirzhnits and A. Linde, Phys. Lett. {\bf 72B}, 471 (1992);
                Ann. Phys. {\bf 101}, 195 (1976)
\bibitem{Jackiw} L. Dolan and R. Jackiw, Phys. Rev. {\bf D 9}, 2904 (1974)
\bibitem{Rubakov} V. Kuzmin, V. Rubakov, and M. Shaposhnikov, Phys. Lett.
               {\rm 155B}, 36 (1985)
\bibitem{McLerran} P. Arnold and L. McLerran, Phys. Rev. {\bf D 36},
              581 (1987)
\bibitem{Klink} F. Klinkhammer and N. Manton, Phys. Rev. {\bf D 30},
              2212 (1984)
\bibitem{Carson} L. Carson, X. Li, L. McLerran, and R. Wang, Phys.
              Rev. {\bf D 42}, 2127 (1990)
\bibitem{Junker} J. Baacke and S. Junker, Phys. Rev. {\bf D 49}, 2055 (1994)
\bibitem{Landsman} N. Landsman, Nuc. Phys. {\bf B 322}, 498 (1989)
\bibitem{FKRS} F. Farakos, K. Kajantie, K. Rum\-mu\-kain\-en,
             and M. Sha\-posh\-ni\-kov,
             Nuc. Phys. {\bf B 425}, 67 (1994); Phys. Lett. {\bf B 336},
             494 (1994); Nuc. Phys. {\bf B442}, 317, 1995
\bibitem{FKLRS} K. Kajantie, M. Laine, K. Rum\-mu\-kain\-en,
            and M. Sha\-posh\-ni\-kov, CERN\_TH/95-226, e-mail archive
	    hep-ph/9508379
\bibitem{DPSSG} D. Diakonov, M. Polyakov, S. Sieber, J. Schaldach, and
             K. Goeke, Phys. Rev. {\bf D 49}, 6864 (1994)
\bibitem{Baacke1} J. Baacke, H. So, and A. Surig, Z. Phys. {\bf C 63},
	    689 (1994)
\bibitem{Baacke2} J. Baacke and A. Surig, preprint DO-TH-95-08,
	    hep-ph/9505435
\bibitem{Patkos} A. Jakovac, K. Kajantie, and A. Patkos, Phys. Rev.
            {\bf D 49}, 6810 (1994)
\bibitem{CarsonI} L. Carson and L. McLerran, Phys. Rev.
              {\bf D 41} , 647 (1990)
\bibitem{Junker2} J. Baacke and S. Junker, Phys. Rev. {\bf D 49}, 4227 (1994)
\bibitem{Shap} A. Bochkarev and M. Shaposhnikov Mod. Phys. Lett. {\bf A 2},
             417 (1987)
\bibitem{Arnold} P. Arnold and O. Espinoza, Phys. Rev. {\bf D 47},
	   3546 (1993)
\bibitem{Dine} M. Dine, R. Leigh, P. Huet, A. Linde, and D. Linde,
            Phys. Rev. {\bf D 46}, 550 (1992)
\bibitem{Farrar} G. Farrar and M. Shaposhnikov, Phys. Rev. {\bf D 50},
	   774 (1994)
\bibitem{DPSSG2} D. Diakonov, M. Polyakov, P. Sieber, J. Schaldach, and
	   K. Goeke, RUB-TPII-02-95, hep-ph/9502245
\bibitem{someone} N. Turok and J. Zadrozny, Nucl. Phys. {\bf B 369},
	   729 (1992); A. Davies et. al., Phys. Lett. {\bf B 336}, 464 (1994)
\bibitem{DPY} D. Diakonov, V. Petrov, and A. Yung, Yad. Fiz. {\bf 39},
            240 (1984) [Sov. J. Nuc. Phys. {\bf 39}, 385 (1984)]
\bibitem{Bohm} M. B\"{o}hm, H. Spiesberger, and W. Hollik, Fortsch. Phys.
	   {\bf 34}, 687 (1986)
\end{thebibliography}
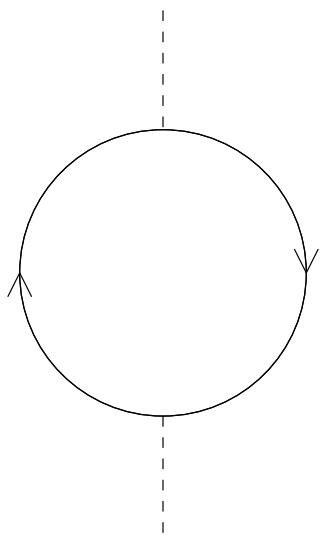
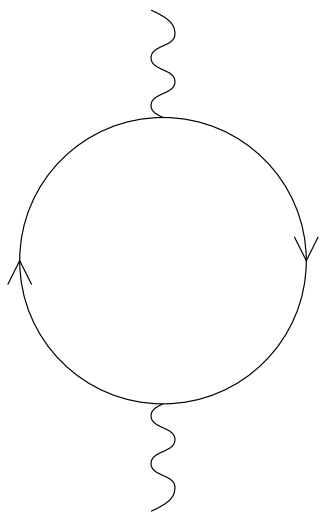
\end{document}